\documentclass[12pt]{article}
\usepackage{cite}
\usepackage{times}
\usepackage{fullpage}
\usepackage{ifthen}
\usepackage{url}
\newcommand{\spacing}[1]{\renewcommand{\baselinestretch}{#1}\large\normalsize}
\spacing{1}

\makeatletter \DeclareRobustCommand\cite{\unskip \@ifnextchar [{\@tempswatrue\@citex}{\@tempswafalse\@citex[]}} \def\@cite#1#2{$^{\hbox{\scriptsize{#1\if[@tempswa , #2\fi}}}$} \makeatother

\renewenvironment{abstract}{%
    \setlength{\parindent}{0in}%
    \setlength{\parskip}{0in}%
    \bfseries%
    }{\par\vspace{-6pt}}

\newcommand{\araa}{Annu. Rev. Astron. Astrophys.}   % Annual Review of Astron and Astrophys
\newcommand{\apj}{Astrophys. J.}   % Astrophysical Journal
\newcommand{\apjl}{Astrophys. J. Lett.}   % Astrophysical Journal, Letters
\newcommand{\apjs}{Astrophys. J. Suppl. Ser.}   % Astrophysical Journal, Supplement
\newcommand{\aap}{Astron. Astrophys.}   % Astronomy and Astrophysics
\newcommand{\mnras}{Mon. Not. R. Astron. Soc.}   % Monthly Notices of the RAS
\newcommand{\pasj}{Publ. Astron. Soc. Jpn}   % Publications of the Astron. Soc. of Japan (note no full stop following Jpn)
\usepackage{color,amsmath,amssymb}
\usepackage{graphicx,units,wasysym,hyperref}

\usepackage{lineno}
\usepackage{caption}

\usepackage[normalem]{ulem}

\usepackage{color}

\makeatletter
\def\@maketitle{
\begin{flushleft}
{\LARGE\bfseries\textsf{\@title} \par}
\end{flushleft}
\begin{flushleft}
{\@author \par}
\end{flushleft}
}

\title{Chlorine and Potassium Enrichment in the Cassiopeia A Supernova Remnant}

\author{XRISM collaboration}
\begin{document}

\date{}

\maketitle

%\begin{affiliations}

%  \end{affiliations}
%%%%%%%%%%%%%%%%%%%%%%%%%%%%%%%%%%%%%%%%%%%%%%%%%%%%%%%%%%%%%%%%%%%%%%%%%%%%%%%

\begin{abstract}
The elements in the universe are synthesized primarily in stars and supernovae, where nuclear fusion favors the production of even-Z elements. In contrast, odd-Z elements are less abundant and their yields are highly dependent on detailed stellar physics, making theoretical predictions of their cosmic abundance uncertain. In particular, the origin of odd-Z elements such as phosphorus (P), chlorine (Cl), and potassium (K), which are important for planet formation and life, is poorly understood. While the abundances of these elements in Milky Way stars are close to solar values, supernova explosion models systematically underestimate their production by up to an order of magnitude, indicating that key mechanisms for odd-Z nucleosynthesis are currently missing from theoretical models.
Here we report the observation of P, Cl, and K in the supernova remnant Cassiopeia A using high-resolution X-ray spectroscopy with XRISM, with the detection of K at above the 6$\sigma$ level being the most significant finding. Supernova explosion models of normal massive stars cannot explain the element abundance pattern, especially the high abundances of Cl and K, while models that include stellar rotation, binary interactions or shell mergers agree closely with the observations. Our observations suggest that such stellar activity plays a significant role in supplying these elements to the universe. 
\end{abstract}

Observing the chemical composition of supernovae (SNe) and their remnants (SNRs) provides direct evidence of how elements are synthesized and distributed in the universe. The elemental abundance patterns of these objects or stars reflect the physical processes occurring in stellar interiors and explosions \cite{hughes2000,maasseron2020,xing2023}, offering crucial insights into the origin and evolution of elements. While even-Z elements (i.e., $\alpha$-elements) have been extensively studied in SNe and SNRs, observations of odd-Z elements, such as phosphorus (P), chlorine (Cl), and potassium (K), remain scarce, and their theoretical treatment is incomplete \cite{koo2013,bekki2024}. Notably, galactic chemical evolution models struggle to explain the observed abundances of these elements \cite{timmes1995,kobayashi2006,kobayashi2020,prantzos2018}, and discrepancies with theoretical predictions have also been reported in observations of some metal-poor stars \cite{tominaga2007}. These inconsistencies highlight a fundamental gap in our understanding of nucleosynthesis processes. Currently, stellar processes such as rotation, binary interactions, and shell mergers (the merging of adjacent nuclear-burning shells into a single convective shell) have been proposed as plausible mechanisms to resolve this issue \cite{prantzos2018,ritter2018,farmer2023}. Here, the nucleosynthesis pathways for odd-Z element production, which include $\gamma$-reactions and neutron/proton capture, are complex\cite{ritter2018}, and it is necessary to directly discuss the synthesis yields of these elements based on observations. However, because emission lines from odd-Z elements are faint, it has been difficult to observe them in SNe/SNRs with existing detectors (particularly in X-rays), and their origin therefore remains unresolved.

Recent near-infrared spectroscopic studies of Cassiopeia A, the youngest known core-collapse SNR in our Galaxy, have reported the detection of emission from P \cite{gerardy2001,koo2013}. In particular, its [P/Fe] ratio is up to 100 times higher than the average in the Milky Way \cite{koo2013}, prompting a discussion on the nucleosynthesis processes during the SN explosion. However, the observed [P/Fe] ratios were scattered by about two orders of magnitude in the remnant because the two elements are synthesized at different locations and unevenly mixed in a SN, preventing a clear understanding of their origin. In addition, since most of
the SN ejecta have already been heated by the reverse shock  ($\lesssim 20$\% of ejecta are unshocked) \cite{laming2020}, X-ray emission from the shocked ejecta  most accurately reflects the true elemental composition in the remnant. To fully understand the origin of odd-Z elements, it is thus crucial to systematically measure the abundances of several  odd-Z elements, including Cl and K, together with neighbouring even-Z elements (e.g. Si, S, Ar, Ca) produced at the same site using X-ray spectroscopy. Therefore, high-resolution non-dispersive X-ray spectroscopy, which has the potential to observe all these elements, is a unique tool to probe the nucleosynthesis of odd-Z elements produced in SNe.

\begin{figure}
%\internallinenumbers
\begin{center}
\includegraphics[bb=0 0 1638 1444, width=1.0\linewidth]{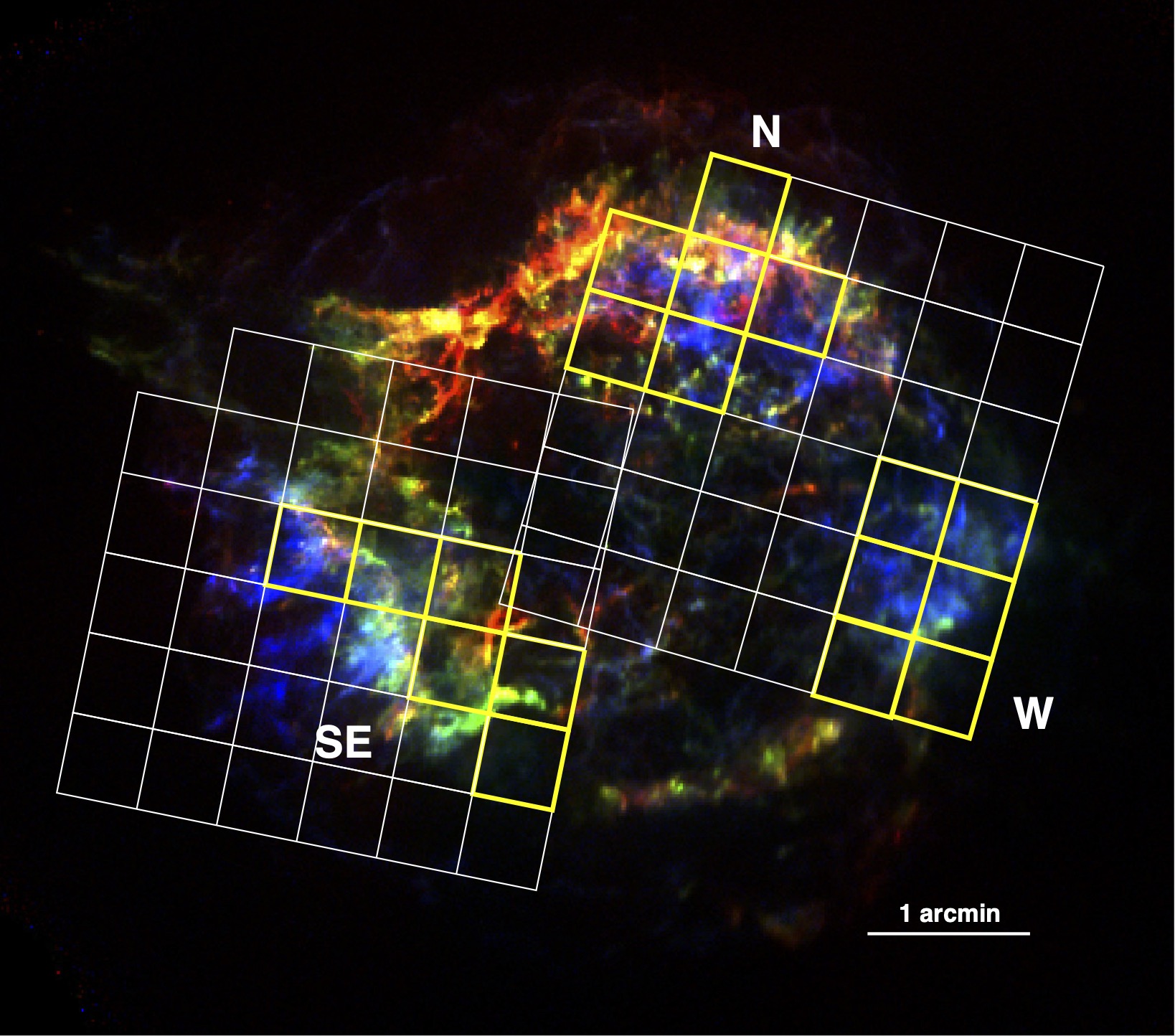}
\end{center}
\caption{\textbf{Three-color image of the core-collapse SNR Cas~A.} Red, green, and blue represent Chandra X-ray images of O- (0.6 to 0.85~keV), Si- (1.76 to 1.94~keV), and Fe-enhanced (6.54 to 6.92~keV) regions, respectively. The white grids indicate the fields of view of our two-pointing observations with XRISM Resolve, which has a $6\times6$ pixel array: a pixel at a corner is only used for calibration. Pixels we selected for our spectral analysis are highlighted in yellow color. In the southeast (SE) and north (N) regions, we examined the spectrum of each pixel and selected those exhibiting large equivalent widths of K emission; the spatial distribution of these pixels closely matches that of the O and Si emission lines. In contrast, the west (W) region exhibits weak O line intensity, and the K emission line in its spectrum is faint (see Figure~\ref{fig:CasA_spec}).}
\label{fig:CasA_image}
\end{figure}

The X-Ray Imaging and Spectroscopy Mission (XRISM) \cite{tashiro2025}, successfully launched on September 7, 2023, is equipped with the Resolve microcalorimeter, capable of non-dispersive high-resolution spectroscopy ($\Delta E<7$~eV) in the X-ray band (1.7--10 keV), where primary emission lines from ionized elements P, Cl, and K are emitted. We conducted observations of Cas~A with XRISM twice in mid-December 2023 (Fig.~\ref{fig:CasA_image} and Table \ref{tab:obs}; see also ``Observation and Data Reduction''). These two pointings cover the majority of the remnant's SN ejecta, with the exception of the northeast-extending jet-like structure, enabling a comprehensive investigation of odd-Z elements in this remnant. Notably, large clumps of ejecta are concentrated in the southeast, north, and west regions, reflecting the explosion's asymmetry \cite{wongwathanarat2017}. As shown in the Chandra elemental maps (Fig.~\ref{fig:CasA_image}), the southeast and north blobs contain O-rich ejecta (which is most likely the product of hydrostatic nuclear fusion in the progenitor 
star's deep interior) along with Fe- and Si-rich ejecta (the products of the explosive nucleosynthesis).
In contrast, the west region is dominated by Fe and Si. We employ spatially-resolved spectroscopy with XRISM/Resolve to examine and compare the elemental abundances across these distinct regions.

\begin{figure}
%\internallinenumbers
\begin{center}
\includegraphics[bb=0 0 3600 3600,width=0.8\linewidth]{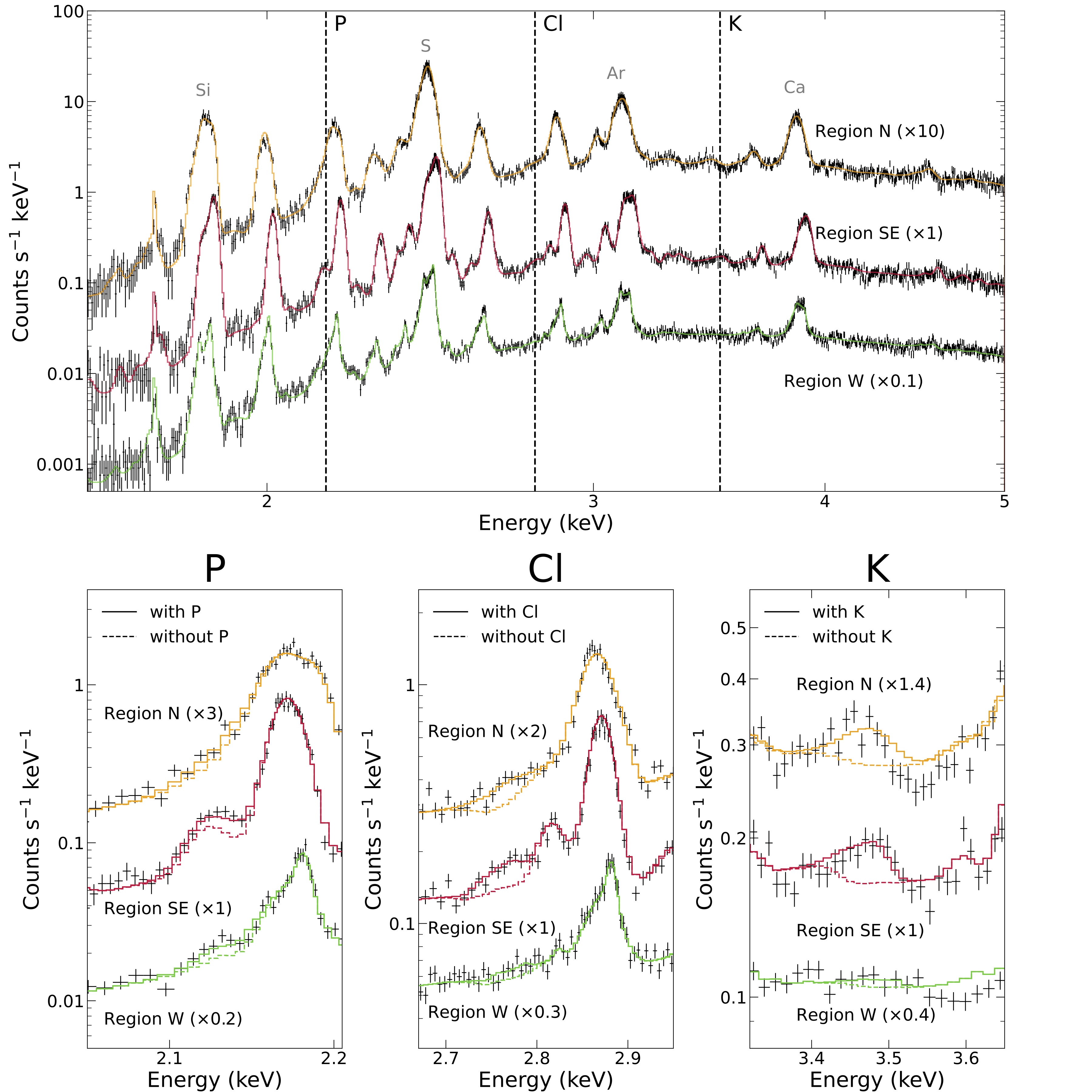}
\end{center}
\caption{
\textbf{Resolve X-ray spectra of Cas~A.} The best-fit models are overplotted as a solid line with colors representing the spectral regions in Figure~\ref{fig:CasA_image} (orange: region N, red: region SE, green: region W). In the top panel, the labels indicate the positions of the He-$\alpha$ emission lines for each element. The vertical broken lines correspond to the centroid energies of the resonance lines of P ($\sim2.152$~keV), Cl ($\sim2.790$~keV), and K ($\sim3.511$~keV) in the rest frame. The bottom three panels show the zoom-in spectra around He-$\alpha$ P (left), Cl (middle), and K (right) line with the best fit models with or without the K emission line. In the spectra of the P band, models without P emission show residuals near 2.13~keV, particularly in the SE and W regions, indicating the signature of P emission in the spectra, while no significant residual is seen in the N region. In the Cl and K bands, models lacking the corresponding lines show residuals in the N and SE regions, but not in the W region. The intermediate-mass elements, particularly in region N, exhibit complex velocity structures (see Figure~4 in Suzuki et al. 2025 \cite{suzuki2025}), which introduces modeling uncertainties. For clarity, the spectrum of region SE has been shifted slightly along the energy axis in the zoomed-in panels.}
\label{fig:CasA_spec}
\end{figure}

Figure~\ref{fig:CasA_spec} shows the X-ray spectra of the north (N), southeastern (SE), and western (W) regions of Cas~A. In the regions SE and N, Resolve clearly detected the emission from highly-ionized He-like ions of Cl and K (at $\gtrsim 5 \sigma$ and $>6 \sigma$ confidence levels, respectively). In contrast, the W region shows no clear line structures and only marginal Cl- and K-line detection significances ($<2\sigma$). Marginal evidence of P are also detected in the region SE and W ($\gtrsim 4 \sigma$ confidence level; see Table \ref{tab:best-fit} and Extended Data Table~1). These results reveal a spatially inhomogeneous distribution of odd-Z elements in Cas~A, with their emission confined to the O-rich regions in the north and southeast, and absent in the western region. These detections are unaffected by background emission, spatial-spectral mixing from the telescope's point spread function, fitting-range choices, or other factors (see Methods and Extended Data Table~1). The abundance ratios relative to hydrogen are approximately 3--5 solar values, confirming that these elements originate from the SN ejecta rather than the interstellar medium (ISM). Dielectronic recombination and inner-shell excitation lines from other elements lie close to the P and Cl line features, with the former being almost completely overlapped and the latter located immediately adjacent, making accurate abundance estimates dependent on careful plasma modeling. In the K band, there are no significant contaminating lines, allowing for a clean separation of the emission feature. Region N shows complex velocity structures in intermediate-mass elements (see Figure~4 in Suzuki et al. 2025 \cite{suzuki2025}), making abundance measurements there less reliable (see ``Methods'' for detail). We therefore adopt the K/Ar and Cl/S ratios from region SE as representative values, with the K/Ar ratio being the most reliable due to its clear detection and minimal contamination, and the Cl/S ratio also considered reasonably robust.  The Cl/S ratio is estimated to be (Cl/S)/(Cl/S)$_{\odot} = 1.0\pm0.1$ and the K/Ar ratio (K/Ar)/(K/Ar)$_{\odot} = 1.3\pm0.2$, based on proto-solar values from \cite{lodders2009}. Here, we use the abundance ratios relative to neighboring even-Z elements in order to examine the characteristics of the synthesis processes in the subsequent discussion. Across a variety of fitting conditions, the derived abundance ratios remain consistent within a $\sim$10--40\% range, which likely reflects the level of systematic uncertainty in our measurements and does not affect our conclusions (see ``Spectral Analysis'' and Extended Data Table~1). 

\begin{figure}
%\internallinenumbers
\begin{center}
\includegraphics[bb=0 0 1920 1200,width=1.0\linewidth]{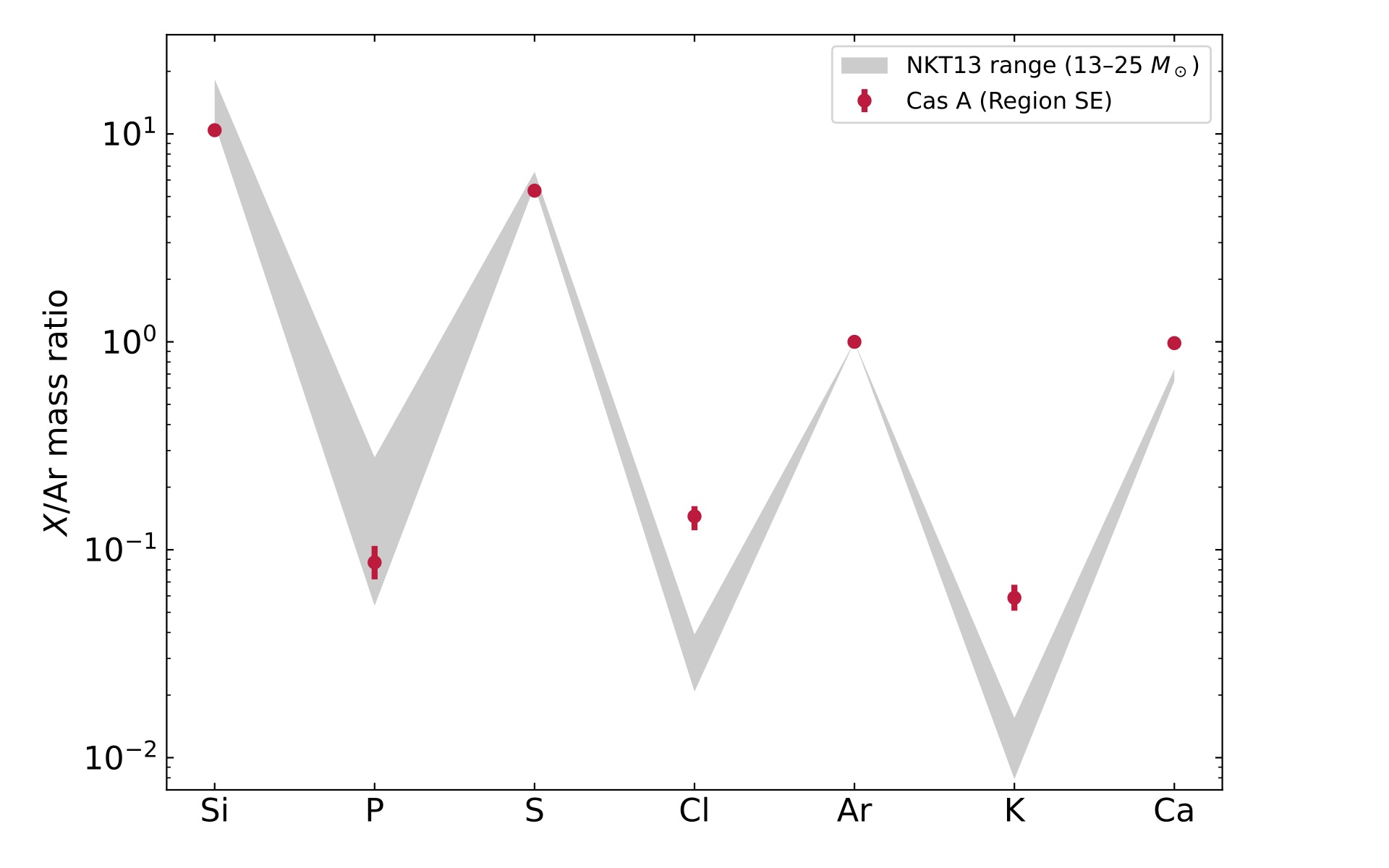}
\end{center}
\caption{\textbf{Elemental mass ratios in Cas~A compared to those in SN nucleosynthesis models.}  
The red circles with error bars represent the mass ratios relative to Ar derived from different regions shown in Figure~\ref{fig:CasA_image}. The error shows a 1$\sigma$ statistical uncertainty. 
The gray region indicates the range predicted by core-collapse SN nucleosynthesis models of Nomoto et al. \cite{nomoto2013}, assuming single-star, non-rotating progenitors with solar metallicity and initial masses in the range 13–25~$M_\odot$. To facilitate comparison between the observed values and the model values, only the model values are connected and that range is shaded. See Extended Data Figure~2 for a more comprehensive comparison across a wider range of model parameters.}
\label{fig:abund}
\end{figure}

The observed ratios close to the solar value suggest an abundant production of Cl and K in the remnant, in contrast to the scarcity in the theoretical models \cite{kobayashi2020}. Figure~\ref{fig:abund} shows the elemental mass ratios from Si to Ca (relative to Ar) obtained from our observation of region SE. It is noteworthy that the observed K and Cl abundances in Cas A, whose initial progenitor mass is estimated to be close to 15~$M_\odot$ \cite{young2006}, are significantly higher than those in SN models whithin the initial mass range 13--25$M_\odot$ from Nomoto et al. (2013) \cite{nomoto2013} (here after NKT13), whereas the observed abundances of the even-Z elements roughly follow the trend of the models (see Extended Data Figure~1 for a more detailed comparison). This implies some mechanisms not included in the models enhanced the abundance of the odd-Z elements in Cas A. The fact that these Cl and K emission are spatially associated with the O-rich ejecta suggests that their production was amplified during the stellar evolution stage, rather than being solely the result of explosive nucleosynthesis. Regarding the P abundance, both observational uncertainties and large variations among theoretical models make quantitative discussion difficult at this stage (see Extended Data Table~1), even though the measured value falls within the range of theoretical predictions in Figure~\ref{fig:abund}.

\begin{figure}
%\internallinenumbers
\begin{center}
\includegraphics[bb=0 0 1800 1400, width=1.0\linewidth]{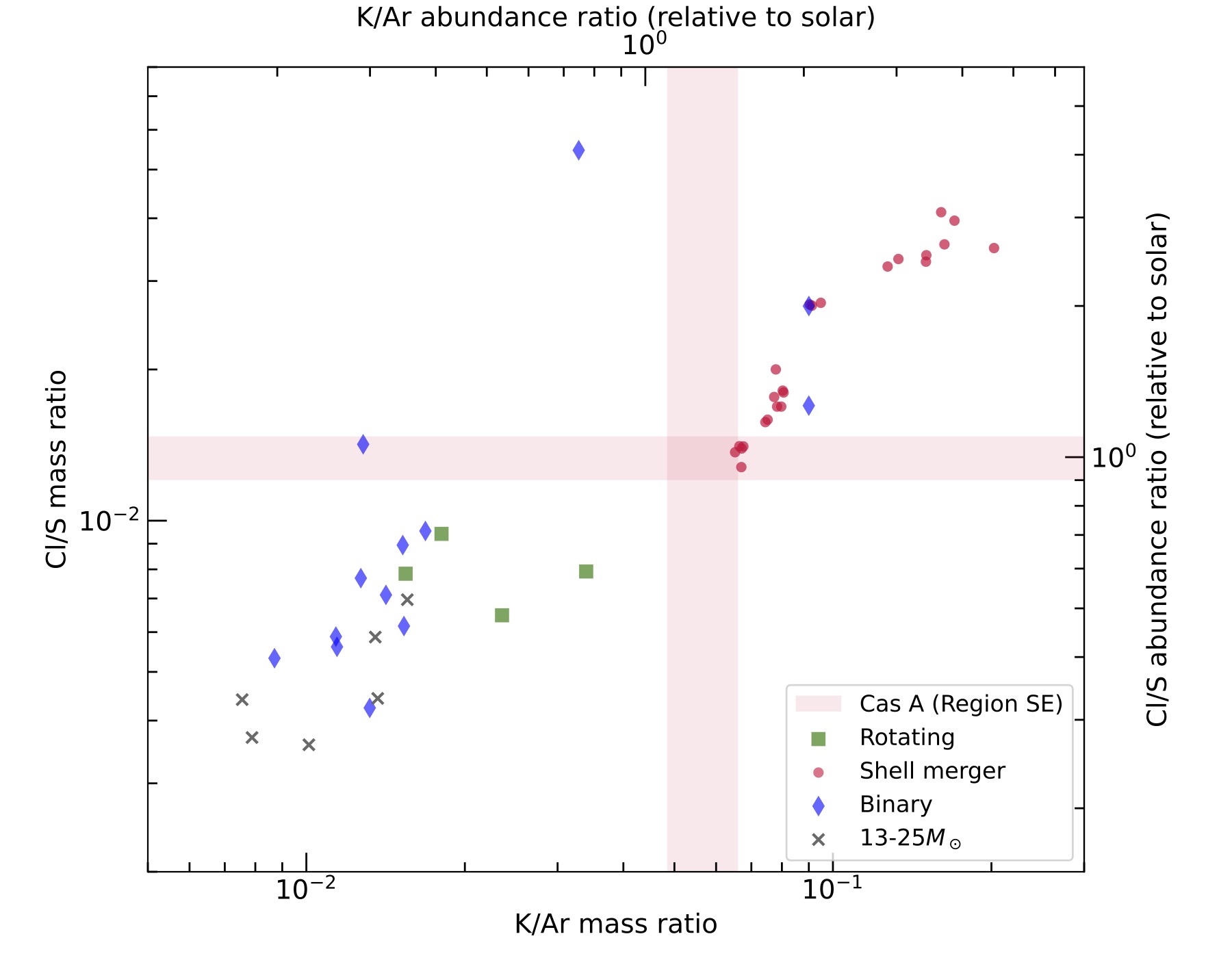}
\end{center}
\caption{
\textbf{Observed Cl/S and K/Ar ratios in Cas~A compared with representative nucleosynthesis models.} The abundance ratio is defined as the elemental abundance ratio normalized to the solar value: (K/Ar)/(K/Ar)$_\odot$ and (Cl/S)/(Cl/S)$_\odot$. The red regions indicate the 1$\sigma$ statistical confidence intervals of observed abundance ratios in region SE. Model predictions are shown for non-rotating single-star progenitors with the initial masses of 13--25$M_\odot$ from Nomoto et al. (2013) \cite{nomoto2013} (gray crosses), rotating models from Limongi \& Chieffi (2018) \cite{limongi2018} with initial masses of 13--25~$M_\odot$ and rotational velocities of 300~km~s$^{-1}$ (green squares), shell merger candidates from a subset of the Sukhbold et al. (2016) \cite{sukhbold2016} models with progenitor masses of $\sim$19--27~$M_\odot$ (red circles), and binary-star models from Farmer et al. \cite{farmer2023} with initial masses of 13--25~$M_\odot$ and companion stars with 80\% of the primary mass (blue diamonds). All models shown assume solar metallicity. See Extended Data Figure~2 for a more comprehensive comparison across a wider range of model parameters.}
\label{fig:cls_kar_main}
\end{figure}

Some stellar activities have been proposed to explain the enhanced production of odd-Z elements in massive stars, most notably stellar rotation \cite{prantzos2018}, binary interactions \cite{farmer2023}, and the shell merger phenomenon \cite{ritter2018}. To investigate these effects, we compare the observed Cl/S and K/Ar ratios with predictions from several core-collapse SN nucleosynthesis models in Fig.~\ref{fig:cls_kar_main}: non-rotating single-star models from NKT13, rotating models from Limongi \& Chieffi (2018) \cite{limongi2018}, shell merger candidates from Sukhbold et al. (2016) \cite{sukhbold2016}, and binary-star models from Farmer et al. (2023) \cite{farmer2023} (see Extended Data Figure~2 for a more detailed comparison). The models with the stellar effects tend to have higher Cl/S and K/Ar ratios than those of the NKT13 models, implying that these processes are required to explain our observations. In particular, some of the shell-merger and binary models predict ratios exceeding those measured in Cas A, indicating an efficient odd-Z production mechanism in these scenarios. 
However, we note that these models involve significant uncertainties. In the case of shell merger models, the predicted yields are highly sensitive to the multidimensional mixing in the burning layer \cite{ritter2018}. Furthermore, Farmer et al. (2023) \cite{farmer2023} point out that, in their binary models, the convective structure near the nucleosynthesis region may introduce substantial uncertainties in the production of these elements. Since our work presents the observational constraint on Cl and K abundances in a core-collapse SN, further theoretical development will be crucial to elucidating their production mechanisms. Detailed comparison and extended discussion of our model analysis are provided in ``Model dependence on odd-Z yields'' of the Methods section. These interpretations are constrained by current model uncertainties, highlighting the need for future efforts to fully integrate observational and theoretical constraints across diverse progenitor scenarios, as well as for population studies combined with theoretical modeling to determine the prevalence of candidate mechanisms among massive star progenitors.

The observational characteristics of the Cas A remnant are generally more consistent with a progenitor that would support these stellar processes than the standard stellar nucleosynthesis. As the remnant of a Type IIb SN, Cas A likely lost most of its hydrogen envelope through binary interaction \cite{krause2008}. Its extremely low neon (Ne) abundance (Ne/O $\approx$ 0.1 solar)\cite{vink1996,hwang2012,sato2025a}, in contrast to the typical Ne/O $\approx$ 1 solar observed in other SNRs \cite{katsuda2010,bhalerao2019}, further suggests unusual stellar processes that deplete Ne. Stellar phenomena such as rotation and shell mergers have been proposed to account for this Ne scarcity in the O-rich layers of the progenitor star \cite{yadav2020, roberti2024, sato2025a}, implying a possible connection between Ne depletion and the enrichment of odd-Z elements during the stellar evolution of Cas~A. In addition, our observation reveals a positive correlation between O emission intensity and odd-Z element abundances, further supporting a stellar nucleosynthetic origin. Although the odd-Z enhancement mechanism remains debated \cite{limongi2018,ritter2018,farmer2023}, this observational link between stellar activity and odd-Z element synthesis provides a unique benchmark. Interestingly, both the O-rich ejecta and odd-Z elements display a biased distribution from southeast to north in the remnant, which may suggest the impact of asymmetric convection in the stellar interior on SN dynamics \cite{mueller2017,bollig2021}. Future observations of the odd-Z elements in Cas~A and other remnants, using XRISM or upcoming X-ray observatories, will provide deeper insights into pre-SN activities and may hold the key to unveiling the chemical history of our galaxy.

\newpage
%%%%%%%%%%%%%%%%%%%%%%%%%%%%%%%%%%%%%%%%%%%%%%%%%%%%%%%%%%%%%%%%%%%%%%%%%%%%%%%

\begin{table}[t]
    \centering
    \caption{Observation log of Cas~A with XRISM.}
    \begin{tabular}{lccc}
    \hline
       Target Name & Sequential No. & Start Date of Obs.  &  Exposure Time (ksec)\\
    \hline \hline
        Cas~A SE & 000129000 & 2023.12.11 05:49:47 & 181.3\\
        Cas~A NW & 000130000 & 2023.12.14 14:38:50 & 166.6\\
    \hline
    \end{tabular}
    \label{tab:obs}
    
    \vspace{0.1cm}
    \noindent\raggedright \textbf{Note.} Sequential No. is the identifier assigned to each observation made by XRISM. Exposure Time is the net duration of the observations after data screening.
\end{table}

\begin{table}[t]
    \centering
    \caption{Best-fit paremeters of spectral fits.}
    \begin{tabular}{lcccc}
    \hline
    Region & \multicolumn{2}{c}{Region SE} & \multicolumn{1}{c}{Region W} & \multicolumn{1}{c}{Region N}\\
       Atomic Database & AtomDB & SPEX & AtomDB & AtomDB\\
    \hline
            (P/Si)/(P/Si)$_\odot$  & $0.9 \pm 0.2$ & $0.9 \pm 0.2$ & $1.7 \pm 0.4$ & $0.8 \pm 0.2$ \\
        (Cl/S)/(Cl/S)$_\odot$  & $1.0 \pm 0.1$ & $0.7^{+0.2}_{-0.1}$ & $0.9 \pm 0.3$ & $0.9 \pm 0.1$ \\
        (K/Ar)/(K/Ar)$_\odot$  & $1.3 \pm 0.2$ & $1.2 \pm 0.2$ & $0.5 \pm 0.3$ & $1.1 \pm 0.2$ \\
        %K/Ar & 1.3$\pm$0.2 & 1.2$\pm$0.2 & 0.5$\pm$0.3 & 1.1$^{+0.2}_{-0.1}$\\
        \hline
        (Mg/H)/(Mg/H)$_\odot$ & 1.9$\pm$0.2 & 2.5$\pm$0.4 & 1.4$\pm$0.3 & 3.1$\pm$0.3\\
        (Al/H)/(Al/H)$_\odot$ & 2.1$\pm$0.5          & 2.1$^{+0.8}_{-0.7}$ & $<$0.7 & $<$0.5\\
        (Si/H)/(Si/H)$_\odot$ & 3.67$\pm$0.04          & 4.19$\pm$0.09 & 1.59$\pm$0.04 & 3.79$^{+0.05}_{-0.09}$\\
        (P/H)/(P/H)$_\odot$ & 3.5$^{+0.7}_{-0.6}$ (5.5$\sigma$)    & 3.6$\pm$0.8 (4.7$\sigma$) & 2.7$^{+0.6}_{-0.7}$ & 3.1$^{+0.7}_{-0.8}$\\
        (S/H)/(S/H)$_\odot$ & 4.04$^{+0.04}_{-0.03}$  & 4.23$\pm$0.09 & 1.85$^{+0.03}_{-0.04}$ & 4.68$^{+0.09}_{-0.11}$\\
        (Cl/H)/(Cl/H)$_\odot$ & 4.2$^{+0.5}_{-0.6}$ (7.7$\sigma$)   & 3.1$^{+0.7}_{-0.6}$ (4.8$\sigma$) & 1.6$^{+0.6}_{-0.5}$ & 4.3$^{+0.7}_{-0.6}$\\
        (Ar/H)/(Ar/H)$_\odot$ & 3.55$\pm$0.06          & 3.74$^{+0.09}_{-0.08}$ & 1.83$^{+0.05}_{-0.06}$ & 4.22$^{+0.08}_{-0.07}$\\
        (K/H)/(K/H)$_\odot$ & 4.5$^{+0.7}_{-0.6}$  (7.0$\sigma$)    & 4.6$^{+0.8}_{-0.7}$ (6.3$\sigma$) & 0.9$^{+0.6}_{-0.5}$ & 4.8$^{+0.8}_{-0.7}$\\
        (Ca/H)/(Ca/H)$_\odot$ & 3.60$\pm$0.07          & 3.69$\pm$0.09 & 2.26$^{+0.07}_{-0.03}$ & 5.14$\pm$0.09\\
        $kT_{e,1}$ [keV]            & 0.92$\pm$0.02 & 0.70$\pm$0.03 & 1.16$\pm$0.03 & 0.61$^{+0.01}_{-0.04}$\\
        $kT_{e,2}$ [keV]            & 2.13$\pm$0.02 & 2.18$\pm$0.06 & 1.78$^{+0.02}_{-0.03}$ & 1.98$^{+0.03}_{-0.01}$\\
        $n_e t_1$ [10$^{10}$ cm$^{-3}$ s] & 3.10$^{+0.10}_{-0.09}$ & 4.9$\pm$0.4 & 28$^{+5}_{-6}$ & 4.0$^{+0.7}_{-0.3}$\\
        $n_e t_2$ [10$^{11}$ cm$^{-3}$ s] & 1.86$^{+0.05}_{-0.03}$ & 1.90$\pm$0.07 & 2.2$^{+0.1}_{-0.2}$ & 1.64$\pm$0.04\\
        Redshift $z_1$ [10$^{-3}$]        & $-$3.8$\pm$0.1 & $-$3.9$\pm$0.2 & 0.53$^{+0.08}_{-0.06}$ & -0.34$\pm$0.2\\
        Redshift $z_2$ [10$^{-3}$]        & $-$4.11$\pm$0.04 & $-$4.11$\pm$0.06 & 4.3$\pm$0.1 & 6.20$^{+0.09}_{-0.04}$\\
        RMS $V_1$ [km s$^{-1}$]           & 1380$\pm$30 & 1370$\pm$40 & 480$^{+30}_{-20}$ & 1030$^{+40}_{-30}$\\
        RMS $V_2$ [km s$^{-1}$]           & 1340$\pm$10 & 1340$\pm$10  & 1830$\pm$30 & 1920$\pm$20\\
        C-value/d.o.f & 7438.28/6777& 1613.18/1248 & 7053.81/6780 & 7208.06/6777\\
    \hline
    \end{tabular}
    \label{tab:best-fit}
    
    \vspace{0.1cm}
    \noindent\raggedright \textbf{Note.} The elemental ratios show elemental abundance ratios relative to the solar values. The error represents a 1$\sigma$ statistical uncertainty.
\end{table}

%%%%%%%%%%%%%%%%%%%%%%%%%%%%%%%%%%%%%%%%%%%%%%%%%%%%%%%%%%%%%%%%%%%%%%%%%%%%%%%

\clearpage

\noindent{\Large\bf Methods} \vspace{0.4cm}\\
%%%%%%%%%%%%%%%%%%%%%%%%%%%%%%%%%%%%%%%%%%%%%%%%%%%%%%%%%%%%%%%%%%%%%%%%%%%%%%%
\paragraph{Observation and Data Reduction:}
We carried out a two-pointing observation of Cassiopeia~A (Cas~A) during the XRISM commissioning phase as summarized in Table~\ref{tab:obs} \cite{suzuki2025,bamba2025,vink2025}.
The Resolve fields of view (FoV) are shown in Figure~\ref{fig:CasA_image}.
Since the Resolve aperture door (``gate valve'') has not opened yet, a 250-$\mu$m-thickness beryllium filter attenuates X-ray photons in the soft X-ray band, reducing the effective area.
The available energy bandpass in our analysis is above approximately 1.6~keV.
The data reduction was done by using calibration data archived in the HEASARC calibration database (CALDB).
Cleaned event data were obtained using the latest release version of the HEASARC software (version 6.34) with the standard screening during post-pipeline processing.
We only used the highest energy resolution (``Hp'') primary events  for the spectral analysis. 
The redistribution matrix file (RMF) was generated with the extra large size option using \texttt{rslmkrmf}. 
The ancillary response file (ARF) was created using \texttt{xaarfgen}, assuming the surface brightness of Cas~A as derived from a Chandra X-ray image in the 2.0--8.0 keV band.

\paragraph{Energy scale reconstruction:}
The gain of the XRISM/Resolve calorimeter detectors is intrinsically unstable since the thermal detectors respond to their thermal environment. The XRISM observatory includes multiple on-board calibration sources to track the time-dependent gain of the Resolve instrument and reconstruct the energy scale for each observation. The Cas A observations were performed early in the performance and verification phase of the mission while the gain reconstruction strategy was still being formalized. Currently, gain fiducials are acquired sparsely, approximately twice per day on average, to sample the slow drift of the detector gain due to changes in spacecraft attitude and the thermal recovery from recycling the sub-kelvin adiabatic demagnetization refrigerator. However, the Cas A observations were performed with much more frequent sampling of the gain: once every other orbit, or approximately every 3 hours, during earth occultation. One position of the instrument's filter wheel (FW) contains a set of $^{55}$Fe radioactive sources providing a fiducial x-ray line at 5.9 keV. During earth occultation, the  FW sources were rotated into the aperture providing ~30 minute exposures with ~500 counts in the 5.9 keV line per pixel, per fiducial measurement. The fiducial measurements were then used to reconstruct the energy scale at that time step using the non-linear reconstruction method described in \cite{porter16}. The energy scale was linearly interpolated between fiducial steps which has been shown to be sufficient to reconstruct the energy scale to better than 0.3 eV at 5.9 keV \cite{porter24}, especially for the frequently sampled fiducials during this early observation. The summed, reconstructed fiducial measurements for the Cas A observations give a line shift error of $<$0.04 eV at 5.9 keV for the entire detector array. More detail can be found in appendix 1 of XRISM collaboration et al. \cite{xrism24}.

\paragraph{Spectral Analysis:}
As shown in Figure~\ref{fig:CasA_image}, we divided the FoV into three regions to investigate the spatial variations of odd-Z elements in the spectral analysis described below. For region SE, we checked the spectrum of each pixel and selected pixels with a large equivalent width of K emissions (pixel ids: 17, 18, 25, 32, 33 and 34), where the distribution of the selected pixels is in good agreement with that of the O emissions.
Each spectral fit was performed using the SPEX software \cite{arnaud1996} version 3.08.01 \cite{kaastra1996} and the Xspec software \cite{arnaud1996} version 12.14.1 (AtomDB 3.1.0.v7) with the maximum likelihood \textit{C}-statistic \cite{cash1979}: we analyzed the Resolve spectrum in the 1.6--5.0~keV band, where any background component is negligible (but, of course we considered the non-X-ray background emissions in our analysis as described below). The spectral fit used in the main text is the result with Xspec. To evaluate the effect of the Non-X-ray background (NXB), we use a temporal NXB spectral model provided by the XRISM calibration team. The NXB model was constructed from a stacked NXB event file based on night-earth observations with a total exposure of 785 ks, and has a power law and 17 Gaussian lines for Al-K$\alpha$1/K$\alpha$2, Au-M$\alpha$1, Cr-K$\alpha$1/K$\alpha$2, Mn-K$\alpha$1/K$\alpha$2, Fe-K$\alpha$1/K$\alpha$2, Ni-K$\alpha$1/K$\alpha$2, Cu K$\alpha$1/K$\alpha$2, Au-L$\alpha$1/L$\alpha$2, and Au-L$\beta$1/L$\beta$2. In Figure \ref{fig:CasA_spec}, the NXB contribution is taken into account in the modeling but is below the displayed range. The NXB model we used has no line features around the K He-like K$\alpha$ emission. We have confirmed that the K/Ar ratios agree within the statistical error (only a few percent change), whether or not the NXB model is considered.

On the basis of the previous observations \cite{hwang2012}, the spectra were fitted with a two-component non-equilibrium ionization (NEI) model absorbed by the interstellar medium (ISM) with the solar abundances \cite{lodders2009}. For the ejecta component, we use a plane-parallel shock (pshock) plasma model (in the case of SPEX, `Mode=3' in the neij model was used). The pshock and single NEI models give consistent results with respect to K detection, although the former gives a slightly better fit (Extended Data Table~1).
We allowed the volume emission measure $\rm{VEM}_{\rm{NEI}}$ $(=n_{\rm{e}} n_{\rm{p}} V_{\mathrm{NEI}}$, where $n_{\rm{e}}$, $n_{\rm{p}}$, and $V_{\mathrm{NEI}}$ are the electron density, proton density, and the volume of the plasma, respectively), the electron temperature $kT_{\rm{e}}$, and the ionization timescale $n_{\rm{e}}t$ to vary.
The abundances of Si, P, S, Cl, Ar, K, and Ca were left free and tied between the two NEI components, while the others were fixed at the solar values. The absorption column density $N_{\rm{H}}$ was fixed at $1.9\times10^{22}$~cm$^{-2}$, $2.0\times10^{22}$~cm$^{-2}$ and $1.5\times10^{22}$~cm$^{-2}$ for regions SE, W and N, respectively, based on the previous observation \cite{hwang2012}. 
Supplementary Figure~1 shows the spectrum of region SE along with the best-fit model consisting of two plasma components. These components, having different electron temperatures and line-of-sight velocities, both contribute to the observed line structures, complicating abundance measurements. For example, region N exhibits a complex velocity structure, with redshifts varying across elements and ionization states \cite{suzuki2025}, which makes it difficult to accurately determine the abundances of Cl and K, even though their spectral features are clearly present. In contrast, region SE shows simpler velocity structure and narrower line widths compared to region N (Figure~\ref{fig:CasA_spec}), allowing for more reliable abundance measurements. We confirmed that additional components such as non-thermal emission or another NEI model do not significantly affect the results for region SE. On the other hand, in regions N and W, if non-thermal emission is not considered, the temperature of the high-temperature plasma component exceeds 5 keV, which is higher than in other regions. Therefore, we added a power-law (non-thermal) component for the spectral fitting of regions N and W. In these regions, previous studies have reported that non-thermal emission is strong\cite{helder2008,uchiyama2008,sato2018,grefenstette2015}, supporting the validity of adding the power-law component in this study.
All the results are summarized in Table \ref{tab:best-fit}, showing that the two-component NEI model effectively explains all the Resolve spectra with statistically acceptable fits. 

Several factors, such as differences in region selections, fit ranges and atomic codes, can introduce errors in abundance estimation. We here discuss systematic errors in abundance estimation (in particular, the K/Ar ratio, which is the focus of this study) by performing spectral analysis under various conditions. The XRISM calibration team does not recommend the asymmetric region selection used here because of the difficulty in evaluating contamination from other regions. For example, it is recommended to select a symmetrical region with a collection of 2$\times$2 pixels. Therefore, we also analyzed the 2$\times$2 pixel region at the center of the observation in the East region (pixel ids; 0, 17, 18, 35), where the odd-Z elements seem to be concentrated. Even in this region, we found the K/Ar ratio to be 1.4$\pm$0.2 with Xspec, which agrees with the result in region A. Here, the detection of K is 7.7$\sigma$. On the other hand, we emphasize that the effect of spatial-spectral mixing due to the point spread function of the telescope does not change our conclusion. In particular, in the eastern region used in the main text, the observed value of K/Ar $\approx$ 1 solar does not change even when using the Resolve full-array data. Also, whether we assume a single NEI plasma or a multi-temperature plasma, the result remains the same. Therefore, slight differences in region selection or plasma parameters do not change our conclusion. In a 2x2 pixel region we can estimate with the raytracing code ``xrtraytrace'' that $\sim$50--60\% of the radiation comes from within this region. Even though there is such a significant amount of contamination from the neighboring regions, it would be difficult to change the plasma parameters to overturn the conclusion.

In Extended Data Table~1, we summarized the K/Ar (also P/Si and Cl/S) ratios in different fitting conditions. It is found that the best-fit values agree within 10\% in most cases. We checked the spectral fitting using a narrower energy band of 3.0--4.2 keV, which agrees well with the results in 1.6--5.0 keV. The electron loss continuum considered in the XL response matrix mainly affects the continuum component below 1.8 keV, so the L matrix was used for the fitting in the 3.0-4.2 keV band without its effect. As mentioned in the main text, the emission lines of Ar and K can be explained by almost a single plasma model, thus we have fit the spectrum using a single NEI (pshock or nei) model in this energy band. The NXB model was not included in this analysis, because the NXB level is three orders of magnitude lower than the emissions from the remnant. In fitting over a wide energy band, the plasma parameters of K and Ar may be affected by other components and their abundance may change. On the other hand, even with this simple approach in the narrow band, the K/Ar ratios agree very well, supporting the robustness of our K/Ar measurements. In the case of Cl/S, the best-fit values agree within 40\% in most cases, which implies larger systematic errors for the Cl observations. This is influenced by the difficulty in modeling the continuum emission; in particular, using a wider fitting energy range tends to result in a larger Cl/S ratio. A similar trend is also seen in the case of the P/Si ratio. These variations likely reflect the magnitude of systematic uncertainties, but even when taking them into account, our conclusions remain robust.

We also investigated a broader energy band from 1.6 to 12 keV, including the Fe-K emissions. In this energy band, it is difficult to ignore the non-thermal emission even in region SE, so a power-law component was added to the analysis (the NXB model is also included). Here, the photon index of the power-law component was fixed at a typical value of 2.3 \cite{patnaude2009}, because it was difficult to determine. We confirmed that the mean values of K/Ar ratios stay within the 1$\sigma$ confidence range with any photon-index values in the range of 2.3--3.0 even in the cases of regions W and N, where the non-thermal components are more dominant than in region SE.
The addition of a power-law component can reduce the fitted continuum level of the thermal component, thereby resulting in higher abundance values as the line to continuum ratio increases. On the other hand, the K/Ar ratio is almost the same as those for the 1.6--5.0 keV band. The result with AtomDB has a slightly larger K/Ar ratio of 1.5, but the spectral fit around the Fe-K band is worse than with SPEX. Currently, there is a descrepancy between atomic codes in the Fe band, and an update is planned. Since the uncertainty may affect the determination of the plasma parameters of K and Ar, the results in the 1.6--5 keV band are considered more reliable for the present analysis.

Supplementary Figure~2 shows a comparison of the spectra in 3.0--4.2 keV generated by AtomDB and SPEX. In the energy band where the Ar and K emissions are present, there are some differences between the different atomic codes, but the spectral models are in good agreement. Several updates were made in these atomic databases to explain this observation of Cassiopeia A. In the case of AtomDB, the new data (version 3.1.0) include significant improvements to the Dielectronic Recombination (DR) satellite lines for B-like to H-like ions; significantly improved inner shell lines from excitation for the same, going up to $n=15$; inclusion of the K-beta and K-gamma lines from inner shell excitation; revised fluorescence yields for $n=3$ and above inner shell lines. In the case of SPEX, a new calculation has been performed for Be- and Li-like Si, P, S, Cl, Ar, and K, focusing on a complete set of innershell and dielectronic recombination transitions up to $n=7$. For the dominant S He- and H-like transitions, SPEX now includes an extended calculation up to $n=52$. The high Rydberg series of He-like S lines reproduce better the observed spectrum around 3.2 keV. Since the fits to the 3.2 keV feature may influence the measurement of the K abundance, a Gaussian component has been introduced into the model with AtomDB to compensate the atomic code difference. The two atomic codes currently agree on K/Ar with a difference of only a few percent.

\paragraph{Model dependence on odd-Z yields:}

Nucleosynthetic studies have predicted that the fraction of the odd-Z elements in SN ejecta depends on several stellar factors, for example, stellar rotation \cite{limongi2018}, binary interaction \cite{farmer2023}, intarnal activities \cite{ritter2018} and initial metalicity \cite{limongi2018}. To investigate these effects, we plot the final yields in ejecta in theoretical models with our observational results in Extended Data Figure~1. The top panel shows the models provided by Limongi \& Chieffi \cite{limongi2018}, in which they calculated the SN yield with different rotation velocoties ($v=0, 150, 300$~km~s$^{-1}$) and metalicities ([Fe/H]=$-3, -2, -1, 0$). While the even-Z elements less depend on the parameters, the odd-Z elements' yields vary from the models with different parameters. In particular, K/Ar ratios appear to increase with both higher metallicities and faster rotational velocities. The middle panel shows the models of Farmer et al. \cite{farmer2023}, in which they surveyed effects of binary interactions by stellar simulations with the companions, whose initial masses were set to be 0.8 times the main stars. In the range of the plot, the K/Ar ratios are likely to be higher in the binary models, but the yields of other elements, including even-Z, seem not to have clear dependence on whether the progenitor was a single star or a binary system. The bottom panel shows the models by Sukhbold et al. \cite{sukhbold2016}. We analyzed the mass fraction profiles at the onset of core collapse, and plot the yields of models in which the O-burning layers are merging with outer C-burning layers (since this resembles the characteristics of a shell merger \cite{yadav2020}, hereafter we refer to it as a ``shell merger candidate'') and those in which the distinct borders between the layers are maintained (see below paragraphs, Supplementary Figure~3, and Sato et al.\cite{sato2025b} for more details). As is consistent with Ritter et al. \cite{ritter2018}, the shell merger candidates show higher Cl and K yields than the other their models.

To compare the relative contributions of these effects to the enhancement of odd-Z elements, we constructed a more comprehensive model plot of the K/Ar and Cl/S ratios, using the models of Nomoto et al. (2013)\cite{nomoto2013}, Limongi \& Chieffi (2018)\cite{limongi2018}, Roberti et al. (2024)\cite{roberti2024}, Farmer et al. (2023)\cite{farmer2023} and Sukhbold et al. (2016)\cite{sukhbold2016} (hereafter NKT13, LC18, R24, F23, and S16), incorpolating our observational results (Extended Data Figure~2).  These models were computed using different stellar evolution codes (e.g., KEPLER, FRANEC), each employing distinct nuclear reaction networks and treatments. However, as shown in Extended Data Figure~2, NKT13, LC18, and S16 models excepted for the shell merger candidates show the similar ratios, implying the variation due to code differences appears small compared to the range caused by physical parameters such as rotation, binarity, and shell mergers. The K/Ar ratios and Cl/S ratios roughly show positive correlation and those in shell merger candidates in S16 models and F23 models are likely to be able to much more enhance the odd-Z yields than rotating models. Both the K/Ar and Cl/S ratios in Cas~A are close to the shell merger candidates in S16 models and both single and binary models of F23, rather than LC18 models incorpolating stellar rotation and metalicity effects. R24 is an extension of LC18 to higher metallicity ([Fe/H] = 0.3, i.e., twice solar). Although increased metallicity is known to enhance odd-Z element production, even the R24 models cannot reach the Cl/S and K/Ar ratios observed in Cas~A. Given that the initial metallicity of Cas~A is suggested to be sub-solar \cite{sato2020}, the contribution of metallicity to the nucleosynthesis of odd‐Z elements in Cas~A is likely to be limited. While the S16 results may suggest that progenitors with initial masses of $\gtrsim$20~$M_\odot$ tend to have enhanced odd-Z yields, it is important to note that the LC18 and NKT13 models in the similar mass range do not show such enhancements. This implies that the enhanced odd-Z element yields are unlikely to be a simple function of initial mass and instead highlight the importance of stellar processes such as shell mergers. We also note that although binary progenitors have been suggested to enhance odd-Z yields, single-star models show variable values of these ratios, and even they can exhibit significantly enhanced odd-Z yields.

While the enhancement of odd-Z elements such as K and Cl is often attributed to progenitor rotation or binary interaction, a closer inspection of our model survey reveals a more nuanced picture, indicating that additional mechanisms may be at play. To explore this point, we further investigated the mass fraction profiles of the models in Figure~\ref{fig:abund} and the results are shown in Supplementary Figure~3. The top panel shows the 14.9$M_\odot$ model from Sukhbold et al. (S16), which retains a textbook onion-like structure. In contrast, the bottom panel exhibits a structure where the Si-rich layer (O-burning layer) appears to be merging with the ONe-rich layer (C-burning layer). The mass fraction profiles of the binary and rotating progenitor models deviate from the canonical onion-skin structure, exhibiting signatures of partial mixing or merging between adjacent burning layers. It has been suggested that such layer merging is more likely to occur in rotating progenitor models \cite{roberti2024}, and some of the binary models used in this study have also been reported to exhibit similar structural features \cite{laplace2021}. Furthermore, in Farmer et al. \cite{farmer2023}, they found that their models with abundant K produced the amount durning their pre-SN phase and pointed out that the convective structure outside the Fe core is the key to the production. These results suggest that the internal structure at the final stage of stellar evolution may influence the yields of odd-Z elements as much as, or even more than, the initial stellar parameters. According to Ritter et al. (2018)\cite{ritter2018}, the pre-SN yields of odd-Z elements in such progenitors depend on the entrainment rate at the O/Si-burning shell. In other words, they are sensitive to the nature of the mixing processes occurring during the O/Si-burning phase shortly before or at the onset of core collapse. Furthermore, some multi-dimensional SN simulations suggest that such a convective state can result in the asymmetry of the remnant \cite{couch2013}. Therefore, future deep observations of the asymmetric SNR Cas A, including spatially resolved abundance distributions, will provide further insight into the nucleosynthetic processes responsible for the excess production of odd-Z elements.

\clearpage

\section*{Data availability}
The XRISM data we used are now publicly available in the archives: \url{https://heasarc.gsfc.nasa.gov/docs/xrism/archive/}.

\section*{Code availability}
To analyse X-ray data with XRISM, we used public software, HEASoft (\url{https://heasarc.gsfc.nasa.gov/docs/software/heasoft/}). We used public atomic data in AtomDB (\url{http://www.atomdb.org/}) and SPEX (\url{https://www.sron.nl/astrophysics-spex}). We fitted the X-ray spectra with a public package, Xspec (\url{https://heasarc.gsfc.nasa.gov/xanadu/xspec/}). 

\section*{Acknowledgments}
This work was supported by JSPS KAKENHI grant numbers JP22H00158, JP22H01268, JP22K03624, JP23H04899, JP21K13963, JP24K00638, JP24K17105, JP21K13958, JP21H01095, JP23K20850, JP24H00253, JP21K03615, JP24K00677, JP20K14491, JP23H00151, JP19K21884, JP20H01947, JP20KK0071, JP23K20239, JP24K00672, JP24K17104, JP24K17093, JP20K04009, JP21H04493, JP20H01946, JP23K13154, JP23K13128, JP19K14762, JP20H05857, JP23K03459, and JP24KJ1485, and NASA grant numbers 80NSSC23K0650, 80NSSC20K0733, 80NSSC18K0978, 80NSSC20K0883, 80NSSC20K0737, 80NSSC24K0678, 80NSSC18K1684, and 80NNSC22K1922. LC acknowledges support from NSF award 2205918. CD acknowledges support from STFC through grant ST/T000244/1. LG acknowledges financial support from Canadian Space Agency grant 18XARMSTMA. AT and the present research are in part supported by the Kagoshima University postdoctoral research program (KU-DREAM). SY acknowledges support by the RIKEN SPDR Program. IZ acknowledges partial support from the Alfred P. Sloan Foundation through the Sloan Research Fellowship. MS acknowledges the support by the RIKEN Pioneering Project Evolution of Matter in the Universe (r-EMU) and Rikkyo University Special Fund for Research (Rikkyo SFR). NW acknowledges the financial support of the GAČR EXPRO grant No. 21-13491X. Part of this work was performed under the aus- pices of the U.S. Department of Energy by Lawrence Livermore National Laboratory under Contract DE-AC52-07NA27344. The material is based upon work supported by NASA under award number 80GSFC21M0002. This work was supported by the JSPS Core-to-Core Program, JPJSCCA20220002. The material is based on work supported by the Strategic Research Center of Saitama University.

\section*{Author Contributions}
T. S. led the Resolve data and pre-SN/SN model analysis and wrote the manuscript. K. M. analyzed the nucleosynthesis models and the Resolve data and wrote the manuscript. H. U. is one of the authors who conceived this study, led the data analysis and prepared the manuscript. L. Gu performed a spatially-resolved odd-Z search and detailed spectral analysis of the remnant, and led the update of the SPEX database. A. F. updated the AtomDB database for explaining the observed spectra. M. A. analysed different combinations of Resolve pixels for the odd-Z element search. He confirmed that the detection of odd-Z elements is robust to small changes in pixel selection. J. V. \& P. P. wrote the PV phase proposal for observing Cas A, which included the case for odd-Z detections using Resolve. P. Plucinsky also analyzed the data and prepared the manuscript. F. S. P. reviewed the Resolve observations of the remnant as an instrumental expert and wrote the Resolve gain section. S. K. helped to make the discussion about the low Ne/O ratio more robust.  S. F. confirmed the abundance of odd-Z elements in a multidimensional SN model. H. Y., A. B., A. S., K. S., E. B. and R. M. helped to improve the manuscript. The science goals of XRISM were discussed and developed over 7 years by the XRISM Science Team, all members of which are authors of this manuscript. All the instruments were prepared by the joint efforts of the team. The manuscript was subject to an internal collaboration-wide review process. All authors reviewed and approved the final version of the manuscript.

\section*{Competing interests}
The authors declare no competing interests.

\clearpage

%\renewcommand{\tablename}{Extended Data Table}
%\setcounter{table}{0}

%\begin{table}[t]
%    \centering
%    \caption{Abundunce ratios obtained from different fitting conditions.}
%    \begin{tabular}{llcccccc}
%    \hline
%Fitting range & Atomic Database & RMF matrix & P/Si & Cl/S & K/Ar & C-value/d.o.f \\
%    \hline
%    {\bf Region SE}\\
%        1.6--5.0 keV & AtomDB & XL & 0.9$\pm$0.2 & 1.0$\pm$0.1 & 1.3$\pm$0.2 &1.10\\
%        1.6--5.0 keV & SPEX & XL & 0.9$\pm$0.2 & 0.7$^{+0.2}_{-0.1}$ & 1.2$\pm$0.2 & 1.29\\
%        3.0--4.2 keV & AtomDB (pshock)& L & --- & --- & 1.3$\pm$0.2 & 1.05\\
%        3.0--4.2 keV & AtomDB (NEI) & L & --- & --- & 1.2$\pm$0.2 & 1.08\\
%        3.0--4.2 keV & SPEX & L & --- & --- & 1.2$\pm$0.2 & 1.15\\
%        1.6--12.0 keV & AtomDB & XL & 1.5$\pm$0.2 & 1.4$\pm$0.2 & 1.5$\pm$0.2 & 0.95\\
%        1.6--12.0 keV & SPEX & XL & 1.2$\pm$0.2 & 1.1$\pm$0.2 & 1.4$\pm$0.2 & 1.24\\
%    {\bf Region W}\\
%        1.6--5.0 keV & AtomDB & XL & 1.7$\pm$0.4 & 0.9$\pm$0.3 & 0.5$\pm$0.3 &1.04\\
%        1.6--15.0 keV & AtomDB & XL & 1.9$\pm$0.4 & 1.2$\pm$0.3 & 0.8$\pm$0.3 & 0.89\\
%    {\bf Region N}\\
%        1.6--5.0 keV & AtomDB & XL & 0.8$\pm$0.2 & 0.9$\pm$0.1 & 1.1$\pm$0.2 &1.06\\
%        1.6--15.0 keV & AtomDB & XL & 0.8$\pm$0.3 & 1.1$\pm$0.2 & 1.4$\pm$0.2 & 0.84\\
%    \hline
%    \end{tabular}
%    \label{tab:best-fit2}

%    \vspace{0.1cm}
%    \noindent\raggedright \textbf{Note.} The elemental ratios show elemental abundance ratios relative to the solar values, such as (X/Y)/(X/Y)$_\odot$.
%\end{table}

\clearpage

\renewcommand{\figurename}{Extended Data Figure} 
\setcounter{figure}{0}

\begin{figure}
%\internallinenumbers
\begin{center}
\includegraphics[bb=0 0 2399 4500,width=0.51\linewidth]{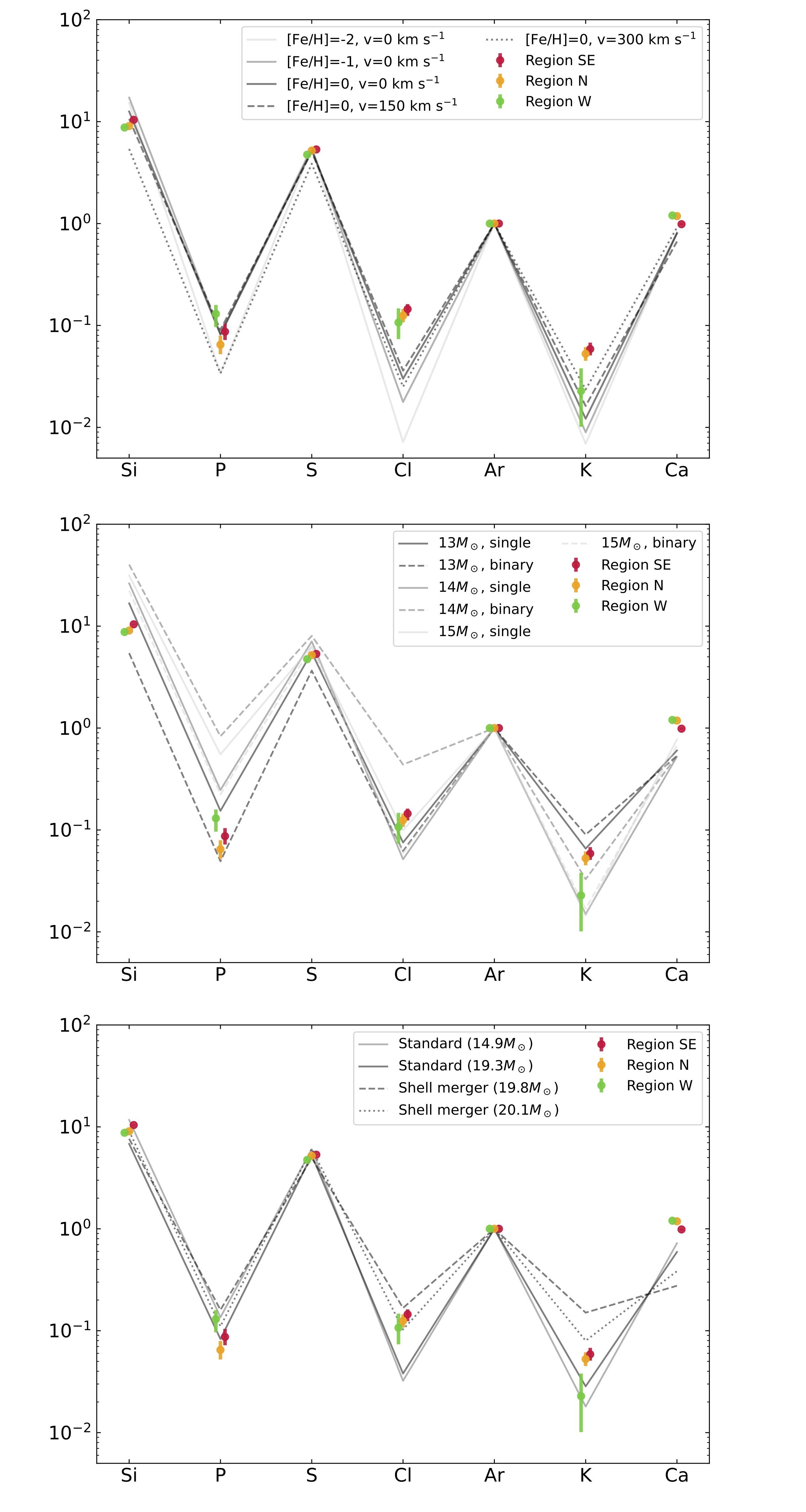}
\end{center}
\caption{\textbf{Elemental mass ratios in Cas A compared to supernova nucleosynthesis models across a wider range of model parameters.} Top: Comparison with 15$M_\odot$ models with the several initial metalicities ([Fe/H]=$-2, -1, 0$) and rotation velocities ($v=0, 150, 300$~km~s$^{-1}$) provided by Limongi \& Chieffi \cite{limongi2018}. Middle: Comparison with 13, 14, 15, 16$M_\odot$ models of the single star cases and the binary cases provided by farmer et al. \cite{farmer2023}. In each binary case, the companion star mass is set to 0.8 times of the progenitor. Bottom: Comparison with non-rotating single star models with several masses provided by Sukhbold et al. \cite{sukhbold2016}. The 14.9$M_\odot$ and 19.3$M_\odot$ models shows onion-like layers constituted with O-, Ne-, and C-burning layers. In contrast, in the 19.8$M_\odot$ and 20.1$M_\odot$ models, the O-burning layers are merging with the outer layers like shell mergers (see Supplementary Figure~3). To facilitate comparison between the observed values and the model values, only the model values are connected with lines.}
    \label{fig:abund_comp}
\end{figure}

\begin{figure}
%\internallinenumbers
\begin{center}
\includegraphics[bb=0 0 2107 1753,width=0.9\linewidth]{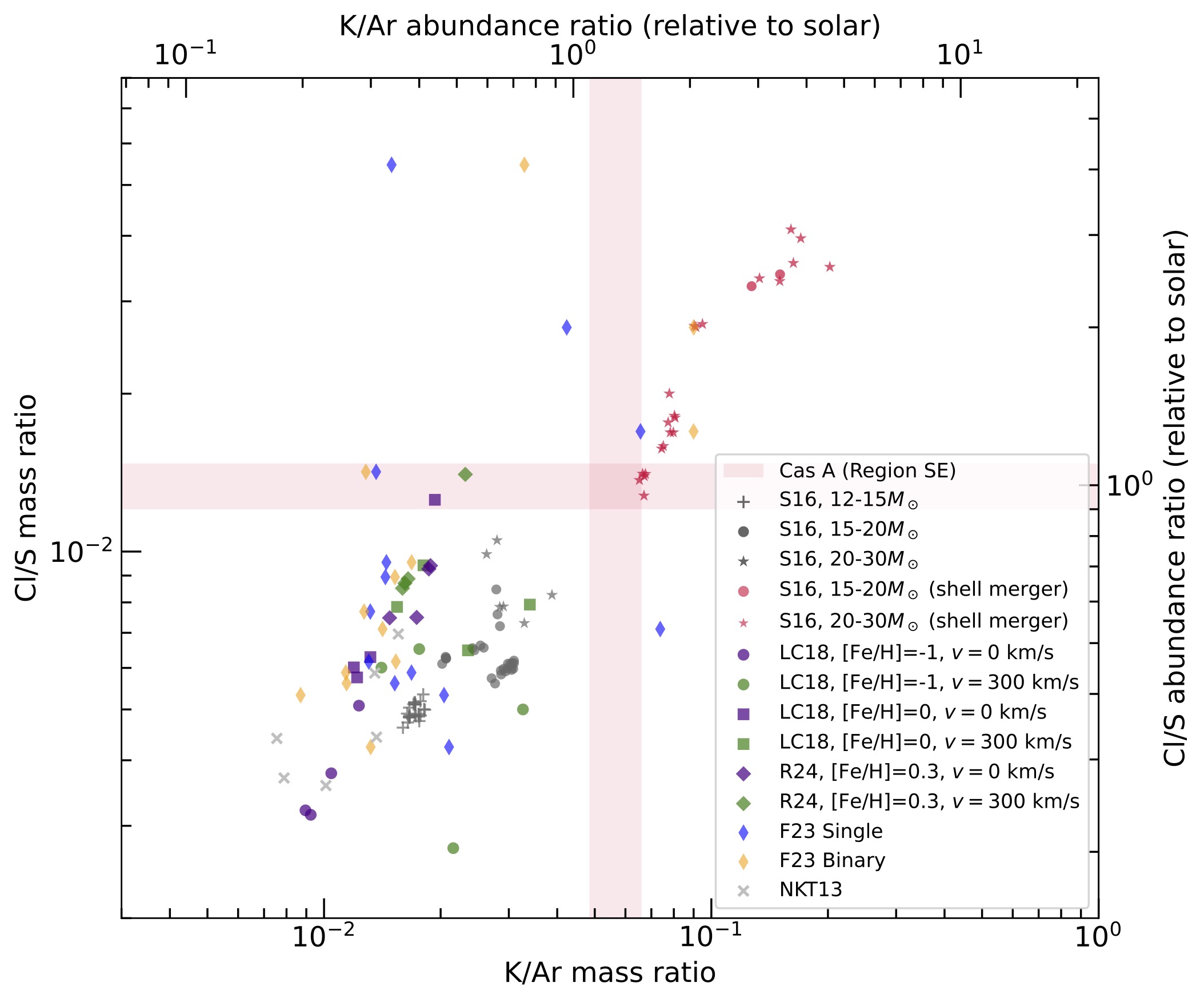}
\end{center}
\caption{
\textbf{Comparison between observed and predicted Cl/S and K/Ar ratios across multiple stellar models.} The hatched light blue regions represent the 1$\sigma$ confidence intervals for the observed mass ratios in region SE of Cas~A. Sukhbold et al. \cite{sukhbold2016} models (12–30~$M_\odot$) are shown with different marker shapes for progenitor mass ranges (circle: 12–15~$M_\odot$, square: 15–20~$M_\odot$, star: 20–30~$M_\odot$). Shell merger candidates are highlighted in red; non-merger models are shown in black. Models from Limongi \& Chieffi \cite{limongi2018} (13–25~$M_\odot$) are shown with squares and circles representing solar and sub-solar metallicity ([Fe/H] = $-1$ and 0), respectively. Rotation is indicated by color: purple for non-rotating and green for rotating models at 300~km~s$^{-1}$. Roberti et al. \cite{roberti2024} models extend LC18 to super-solar metallicity ([Fe/H] = 0.3), and are plotted in a similar format (non-rotating and rotating cases).  Farmer et al. \cite{farmer2023} models (10–30~$M_\odot$) are shown as stars: blue for single-star models and orange for binary models. Nomoto et al. \cite{nomoto2013} models (10–30~$M_\odot$) are indicated as grey crosses. The figure shows that models involving shell mergers or binary interactions can produce enhanced Cl and K abundances, consistent with our observational constraints.}
    \label{fig:cls_kar}
\end{figure}

\clearpage
\section*{Author list}

\noindent
XRISM collabration

\noindent
Marc Audard$^{1}$
Hisamitsu Awaki$^{2}$
Ralf Ballhausen$^{3,4,5}$
Aya Bamba$^{6}$
Ehud Behar$^{7}$
Rozenn Boissay-Malaquin$^{8,4}$
Laura Brenneman$^{9}$
Gregory V.\ Brown$^{10}$
Lia Corrales$^{11}$
Elisa Costantini$^{12}$
% ...

\vspace{0.0cm}
\noindent
% ...
$^{1}$Department of Astronomy, University of Geneva, Versoix CH-1290, Switzerland \\ %Geneva U
$^{2}$Department of Physics, Ehime University, Ehime 790-8577, Japan \\ %Ehime U
$^{3}$Department of Astronomy, University of Maryland, College Park, MD 20742, USA \\ %U of Maryland 
$^{4}$NASA / Goddard Space Flight Center, Greenbelt, MD 20771, USA \\%NASA/GSFC 
$^{5}$Center for Research and Exploration in Space Science and Technology, NASA / GSFC (CRESST II), Greenbelt, MD 20771, USA\\
$^{6}$Department of Physics, University of Tokyo, Tokyo 113-0033, Japan \\% U of Tokyo
$^{7}$Department of Physics, Technion, Technion City, Haifa 3200003, Israel\\
$^{8}$Center for Space Science and Technology, University of Maryland, Baltimore County (UMBC), Baltimore, MD 21250, USA\\
$^{9}$Center for Astrophysics | Harvard-Smithsonian, MA 02138, USA\\ %CfA
$^{10}$Lawrence Livermore National Laboratory, CA 94550, USA\\ %LLNL
$^{11}$Department of Astronomy, University of Michigan, MI 48109, USA\\ %U of Michigan
$^{12}$SRON Netherlands Institute for Space Research, Leiden, The Netherlands\\ %SRON
% ...

\clearpage
\setcounter{page}{1}

\begin{center}
\noindent{\Large Supplementary Information: \\ \vspace{0.1cm} Chlorine and Potassium Enrichment in the
Cassiopeia A Supernova Remnant} \vspace{0.4cm}\\
\end{center}

\renewcommand{\figurename}{Supplementary Figure} 
\setcounter{figure}{0}

\begin{figure}[h]
\begin{center}
\includegraphics[bb=0 0 3250 2369,width=1.0\linewidth]{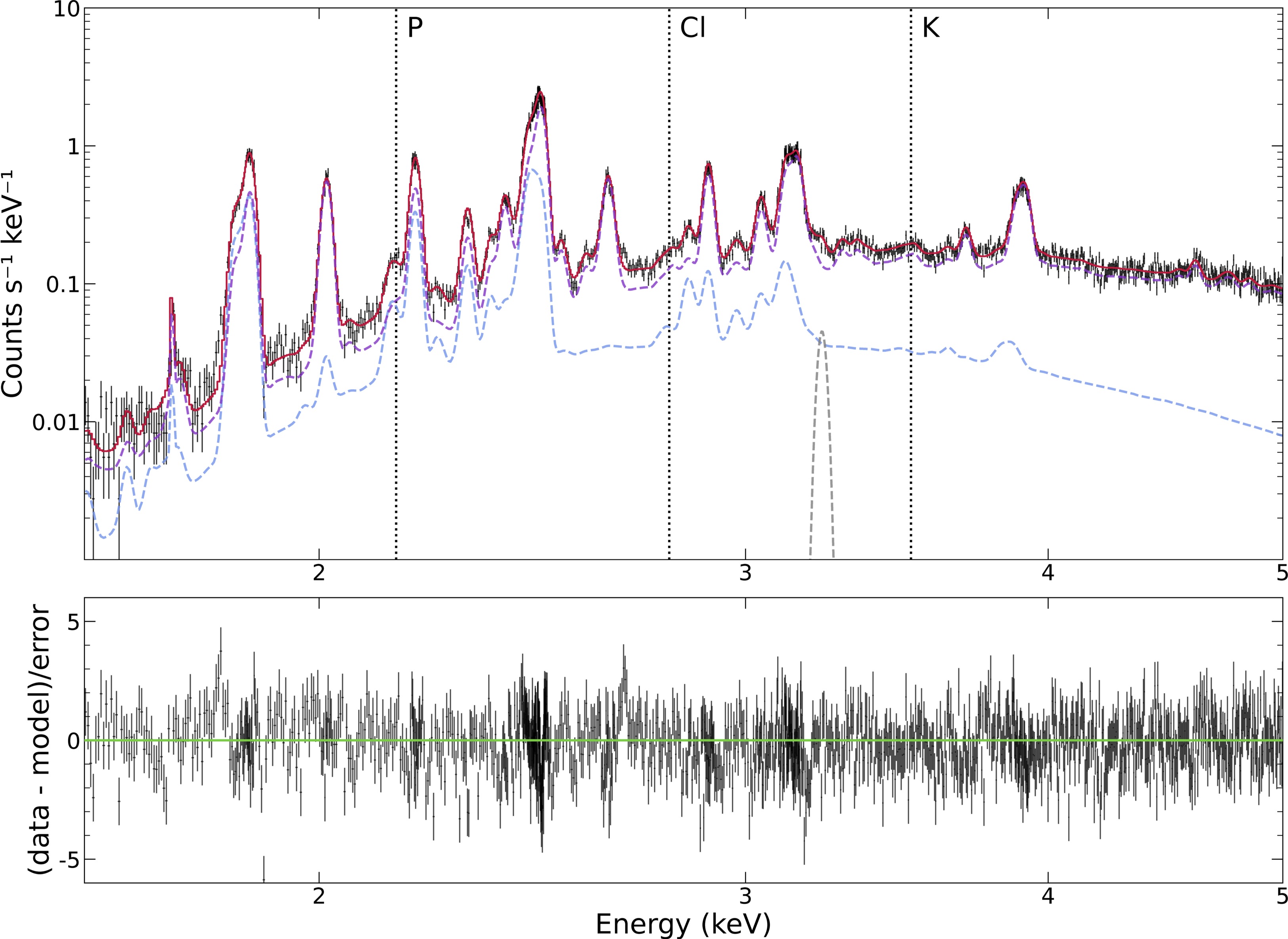}
\end{center}
\caption{\textbf{The Resolve spectrum and the best-fit model of Region SE.} The supectrum is shown as the black points and the total best-fit model of AtomDB is shown as the red line. The purple and blue dashed lines show the two-NEI components, respectively. The grey line shows the Gaussian component adopted to the 3.2~keV structure.}
\label{fig:RegSE_spec}
\end{figure}

\begin{figure}
%\internallinenumbers
\begin{center}
\includegraphics[bb=0 0 2052 811,width=1.0\linewidth]{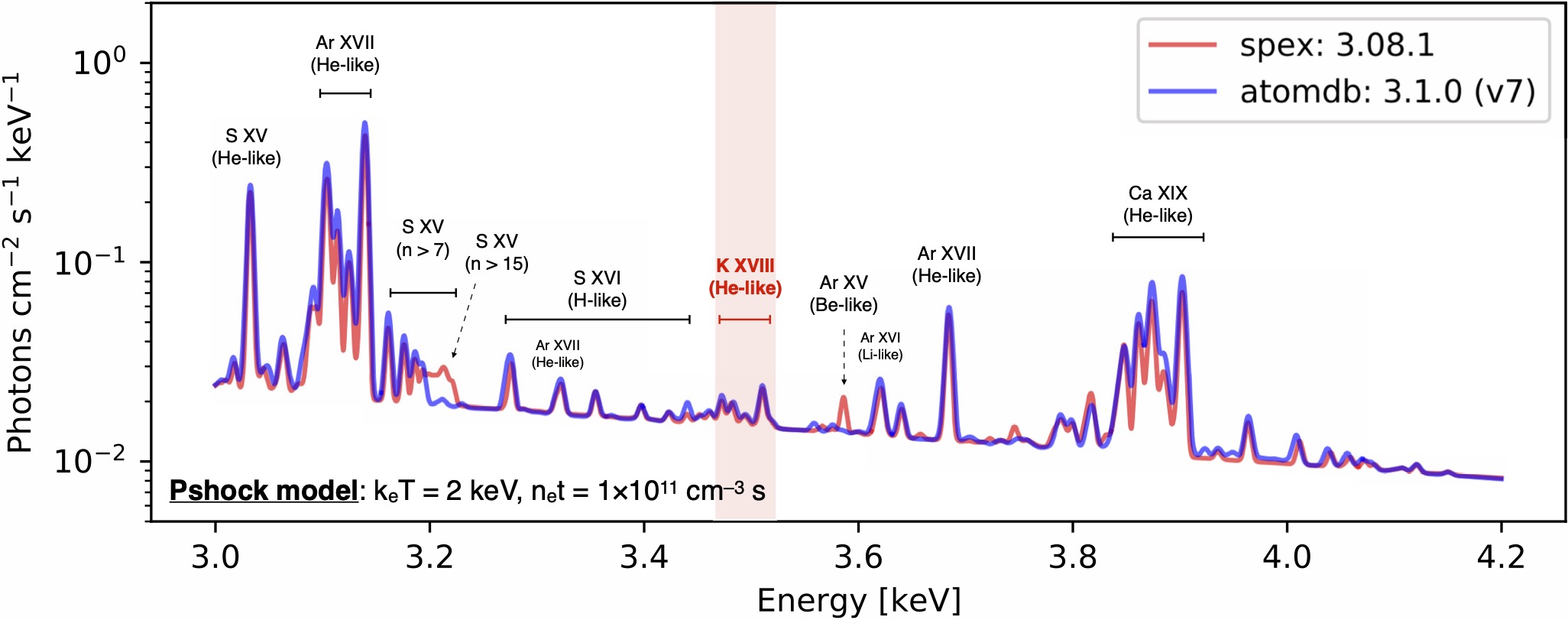}
\end{center}
\caption{\textbf{Comparing different atomic codes}. The red and blue curves show the pshock model for AtomDB (pre-release version 3.1.0.v7) and SPEX (3.08.1). The two models assume the same plasma parameters ($k_e T=2$ keV, $n_e t = 1\times 10^{11}$ cm$^{-3}$ s and RMS $V = 200$ km s$^{-1}$).}
\label{fig:atmcode}
\end{figure}

\begin{figure}
%\internallinenumbers
\begin{center}
\includegraphics[bb=0 0 2000 2800,width=0.8\linewidth]{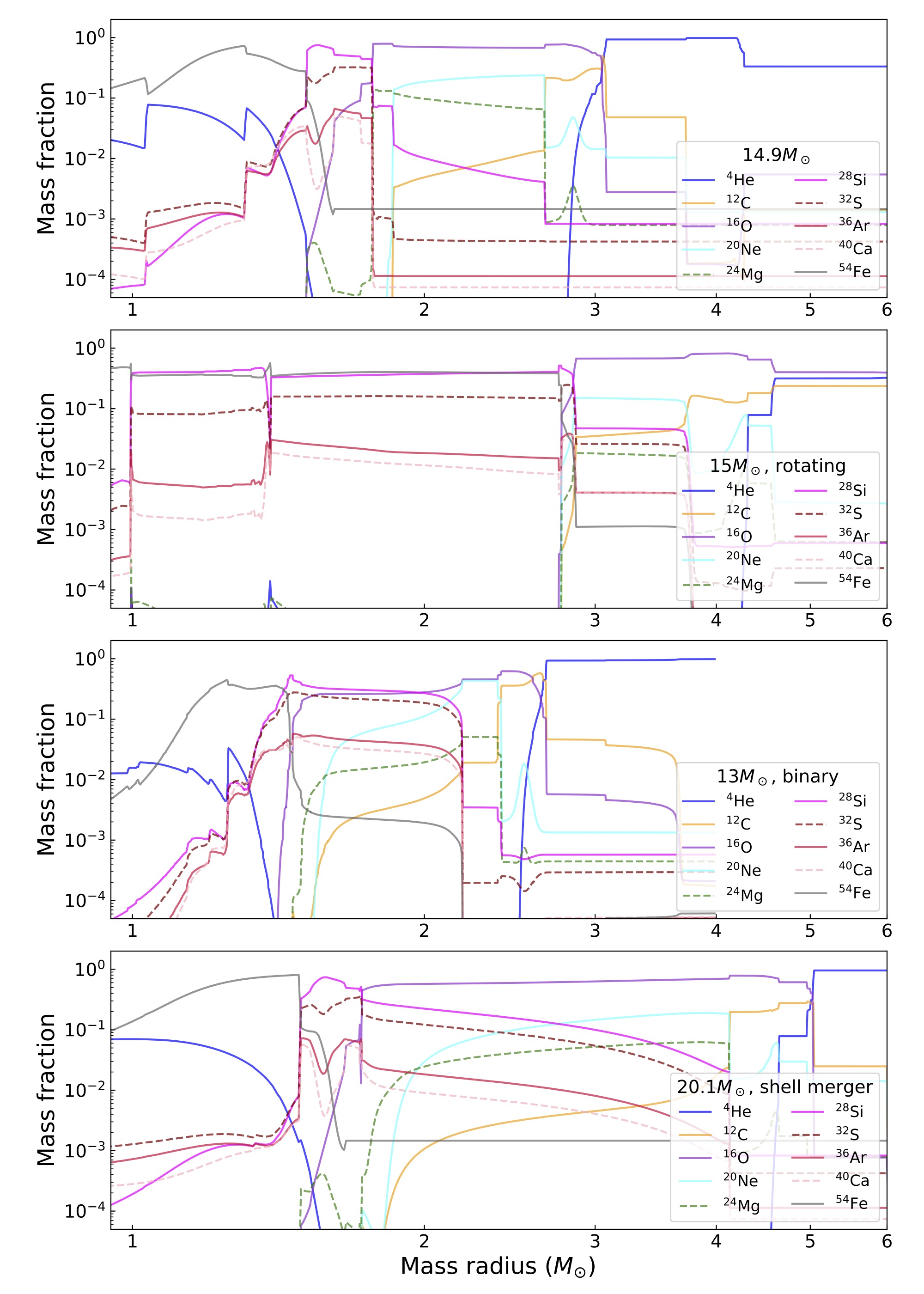}
\end{center}
\caption{\textbf{The mass-fraction profiles of the models in Figure~\ref{fig:abund}.} Top: $14.9M_\odot$ standard model by Sukhbold et al. \cite{sukhbold2016}. Second top: $15M_\odot$ model with an initial rotation velocity of 300~km~s$^{-1}$ by Limongi \& Chieffi \cite{limongi2018}. Third top: $13M_\odot$ progenitor model with a $10.4M_\odot$ companion star by Farmer et al. \cite{farmer2023}. Bottom: $20.1M_\odot$ shell-merger model by Sukhbold et al. \cite{sukhbold2016}.}
    \label{fig:mfrac}
\end{figure}

\end{document}